# Unraveling the Phase-amplitude Coupling Modulation in a Delay-Coupled Diode Lasers Functionality

Pramod Kumar*[,a], Sudeshna Sinha[c], Vishwa Pal[d], and John Gerard McInerney[a, b]

[a] Department of Physics, University College Cork, Ireland;
[b] Optoelectronics Group, Tyndall National Institute, Lee Maltings, Cork, Ireland;
[c] Department of Physics, Indian Institute of Science Education and Research Mohali, Mohali, Punjab, India;
[d] Department of physics of complex networks, Weizmann Institute of Science, Rehovot, Israel.

## ABSTRACT

We numerically investigate the effect of phase-amplitude coupling modulation on power spectra in semiconductor lasers subject to optical injection in a face to face configuration, when a non-negligible injection delay time is taken into account. We find that as phase-amplitude coupling factor α varies, the system goes through a sequence of phase transactions between In-phase locking states to anti-phase locking states via phase-flip bifurcation. Moreover, we observed the signature of frequency discretization (Frequency Island) and uncovered the physical mechanism for the existence of multi-stability near the phase transitions regimes. Within the windows between successive anti-phases to in-phase locking regions, optical injection induced modulation in alpha, unveiling a remarkable universal feature in the various dynamics of coupled lasers system which could be useful in controlling the chirp or pulse repetition rate of a photonic integrated compact device with the aid of phase control.

## 1. INTRODUCTION

Mutually delay-coupled semiconductor lasers (Coherent or Incoherent coupling) have attracted much attenuation in recent years not only of interest from a fundamental point of view but they also play an important role in modern technological applications [1, 2, 3]. Moreover, the excellent controllability of semiconductor lasers in the laboratory make them fascinating test-bed for the experimental investigation of general nonlinear phenomena in the theory of mutually coupled oscillators system [2]. By mutually coherent coupling, we mean the coupled cavity modes (CCM) are phase locked with zero phase difference or fixed relative phase constants between different modes. By mutually incoherent, we mean precisely the opposite: the phase differences between different modes are out-of-phase or anti-phase [2]. Each CCM has a different frequency, but when all are mutually coherent a periodic pulse train is generated, and conversely, when they are not mutually coherent, complex nonlinear dynamics are generated. In a coupled laser system, an amplitude fluctuation in one laser leads to a carrier density fluctuation, and through α, a phase fluctuation in the same laser. A significant change in the relative phase leads to an amplitude change in the second laser and an accompanying change in its carrier density [3, 4]. So, our study clearly indicates the loss of coherence in some groups of modes while other modes remain coherent as the coupling constant (optical injection strength) is varied. We perform detailed numerical computations using a system of delay coupled lasers and confirm that the coexistence of the coherent and incoherent modes may have a fundamental origin resulting from nonlinear strong coupling between the amplitude and the phase of the optical field.

Nonlinear dynamic coupling between phase and amplitude of the optical electric field is caused by carrier-induced variations in real and imaginary parts of refractive index in the laser cavity [1, 2]. Manipulation of phase-amplitude coupling factor (as represented by α) is a major determinant of fundamental aspects of semiconductor lasers, including linewidth, chirp under current modulation, mode stability, and dynamics in presence of optical feedback and injection [5, 6, 7]. In the steady state, α is constant, but varies in highly dynamical regimes or pulse operation where the variation in carrier density is quite large, α has been recently shown to vary with optical feedback and optical injection [5, 7].

*Send correspondence to Pramod Kumar. E-mail: pramod.kumar@ucc.ie

Nevertheless, so far understanding of the impact of optical injection induced modulation in α remains poor. Here, we describe for the first time manipulating α of a semiconductor laser with optical injection in the short cavity leading to frequency discretization and tuning by changing the injection, is demonstrated numerically for the first time.

## 2. THEORETICAL MODEL AND RESULTS

We investigate two mutually delay-coupled lasers (Fig. 1) using dimensionless semi-classical rate equations [1, 2] for the time evolution of the complex electric field $E_j(t)$ and the excess carrier density above threshold $N_j(t)$ averaged

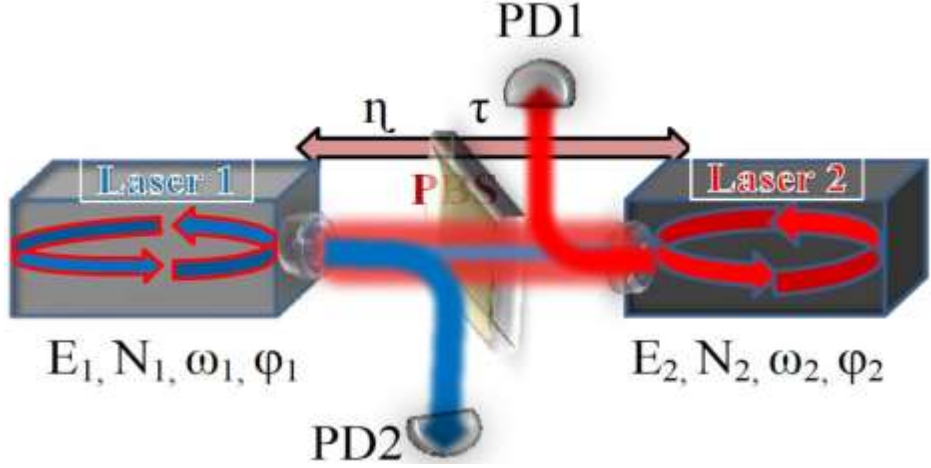

*Figure 1.* . Schematic diagram of a coupled diode laser system: a combination of the polarizing beam-splitter (PBS) and the polarizer (POL) allows a variation of the optical coupling strength η; the photodetectors, PD1 and PD2, measure the output powers P1 and P2 of the respective lasers:

$$\frac{dE_j(t)}{dt} = (1+i\alpha)N_j(t)E_j(t) + \eta E_k(t-\tau)e^{-i\omega_k \tau}, \; j \neq k,$$
$$T\frac{dN_j(t)}{dt} = J_j - N_j(t) - (1+2N_j(t))|E_j(t)|^2, \; j,k = 1,2,$$
(1)

$$\frac{d\phi_1}{dt} = -\alpha N_1(t) - \eta \frac{A_2(t-\tau)}{A_1(t)} \sin[\phi_1(t) - \phi_2(t-\tau) - \omega_2 \tau] \quad (2)$$

$$\frac{d\phi_2}{dt} = -\alpha N_2(t) - \eta \frac{A_1(t-\tau)}{A_2(t)} \sin[\phi_2(t) - \phi_1(t-\tau) - \omega_1 \tau] \quad (3)$$

spatially over the laser medium, the injected excess current densities $J$s above threshold and $P_j = |E_j|^2$. The above equations can be written in terms of field amplitudes using $E_j \sqrt{P_j} e^{-i\phi_j} \equiv A_j e^{-i\phi_j}$ $T$ is the ratio of the carrier lifetime to the photon lifetime, the delay time $\tau = L/c$ is the time taken by the light to cover the distance $L$ between the lasers, and α is the linewidth enhancement factor [6, 7]. The linewidth enhancement factor is the ratio of the change in real component of the complex refractive index with a change in carrier density, to the change in imaginary component of the refractive index with a change in carrier density as

$$\alpha_H = \frac{dn'_p/\partial N}{dn'_z/\partial N} \approx -\frac{4\pi}{\lambda}\frac{\Delta n(t)}{\Delta g(t)} = -\frac{2}{L}\frac{\Delta\phi}{\Delta g} = 20\log(e)\frac{\int_{-\infty}^{\infty}\Delta\phi(t)dt}{\int_{-\infty}^{\infty}\Delta g(t)dt} \quad (4)$$

Where phase change $\Delta\phi = -\frac{2\pi L \Delta n}{\lambda}$, are caused by refractive index changes $\Delta n$ where $\lambda$ is the wavelength of the light.

We start by analyzing the modulation properties of the coupled laser system from the rate equations 1 to 3 and using equation 4. The behavior of the correlation functions versus $\alpha$ shown in figure 2.

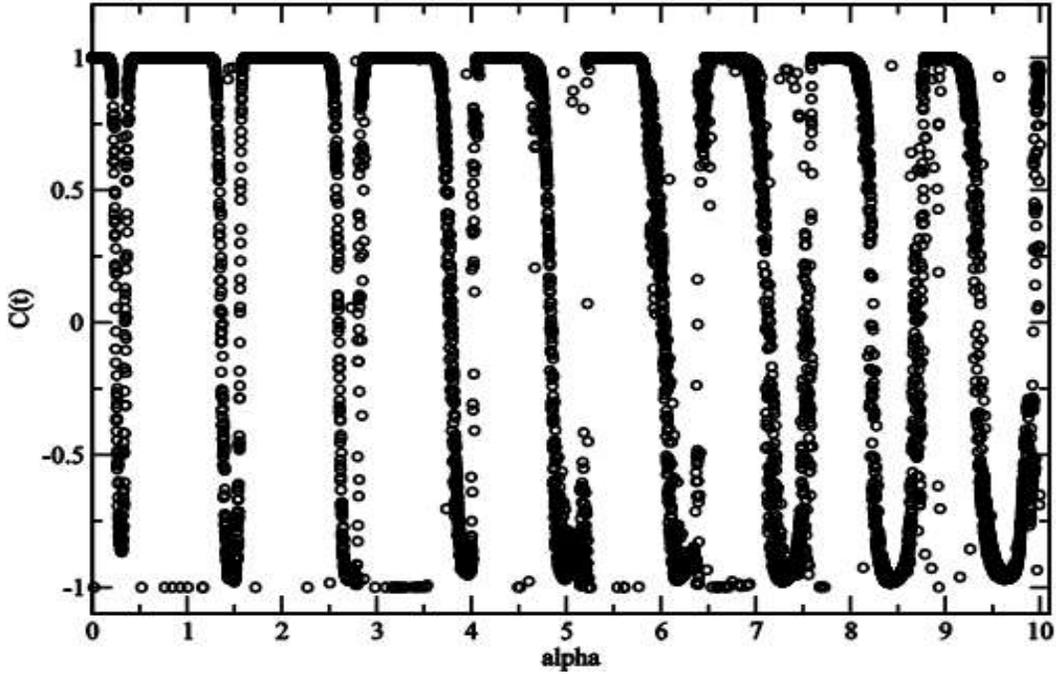

*Figure.* 2. Plot of the cross-correlation C(t) versus $\alpha$ for a fixed time delay $\tau = 14$ (in units of cavity photon lifetime).

In accordance with the α-factor definition using Hakki-Paoli method [11, 12], we can recast the definition of amplitude-phase coupling factor through the following relationship: by tuning the optical injection strength η, by varying the frequency detuning, $\Delta v$ and by varying the injection current, I [8, 9, 10].

$$\alpha_M = -\frac{2\pi}{L\Delta\lambda}\frac{d\lambda/d\eta}{dg/d\eta} \quad (5)$$

$$\alpha_M = -\frac{2\pi}{L\Delta\lambda}\frac{d\lambda/d(\Delta v)}{dg/d(\Delta v)} \quad (6)$$

$$\alpha_M = -\frac{2\pi}{L\Delta\lambda}\frac{d\lambda/dI}{dg/dI} \qquad (7)$$

$$g_{net} = \Lambda_p g - \delta i = \frac{1}{L}\ln[\frac{1}{R}\frac{\sqrt{X-1}}{\sqrt{X+1}}] \qquad (8)$$

We find that in the region where stability of phase-locking switches occur, there exist phase transitions, i.e., phase-locked dynamics can be switched from in-phase locking (antiphase) to antiphase locking (in-phase) and back to in-phase (ant-phase) and so on just by progressive increase of the α for keeping all the control parameter are fixed as shown in figure 2. Specifically shown in figure 3., just by a progressive increase of η, quantization in RF frequencies of phase-locked dynamics changes from being in-phase to anti-phase and back to in-phase and so on for the coupling strength and delay time while RF frequency of the locked dynamics changes from being anti-phase to in-phase and back to antiphase and so on for the another combination of control parameters, coupling strength η and τ.

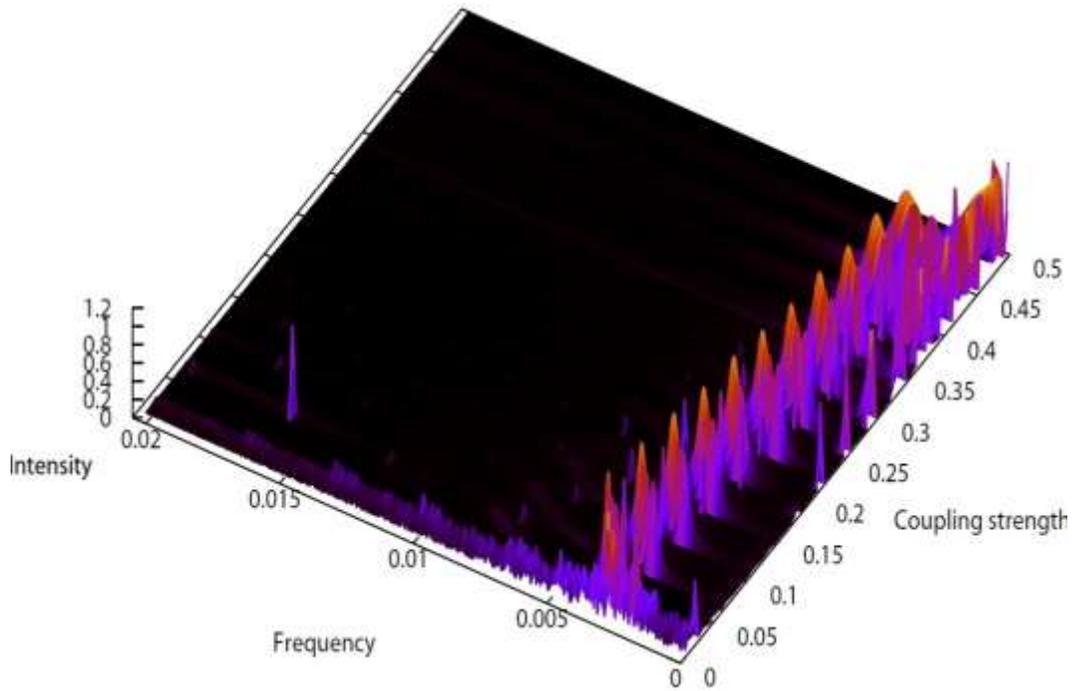

*Figure. 3. Plot of frequency discontinuity vs η for the fixed value of τ=14* (in units of cavity photon lifetime).

In the equation 8 where net modal gain $g_{net}$ and $\Lambda p$ is the optical confinement factor. $X$ is the ratio of the peak-to-valley intensity levels. The refractive index increases (decreases) with the carrier decreases (increases). Consequently, the emission frequency becomes proportional to the carrier number. The relative change in the carrier-dependent resonance around the laser threshold can then be expressed by the following relation [9, 13]

$$\Delta\omega(N) = \frac{\alpha}{2}g\Delta N, \qquad where \ \Delta N \equiv N - N_{th} \qquad (9)$$

Accordingly, the injection-locked laser may operate at a frequency different from its cavity resonance condition, namely operating at resonance frequency and given by the following expressions [9, 13]:

$$\Delta\omega(N) = 2\pi\Delta\nu = \frac{\alpha}{2}g\Delta N \qquad (10)$$

$$\omega(resonance) = \Delta\omega\ (injection) - \frac{\alpha}{2}g\Delta N,$$
$$where\ \Delta\omega(injection) \equiv \omega(injection) - \omega_0 \qquad (11)$$

$$\beta = \beta_{RO} + \frac{K}{2\sqrt{J}}\sqrt{1+\alpha^2}\cos\varphi(t) - \varphi_0\ where\ \beta_{RO} = \gamma\frac{1+2J}{2} \qquad (12)$$

We have modified the RO (Relaxation Oscillations rate) and find an expression for the effective damping β in equation (12) near the phase transitions regimes [9]. Note that this damped oscillation differs from the relaxation oscillation in the physical mechanism, because the relaxation oscillation results from an interaction or coupling between field amplitude (or photon) and carrier through the stimulated emission and, therefore, the cavity resonance is not required to explain such a mechanism ( Eq. 10 to Eq. 12). The coupling strength η are tuned such that the proportionate injection field to each laser is the same [5]. We have observed the multistability in the gap between two consecutive frequency islands after solving the equations (1) and (2) as shown in Figure. 4.

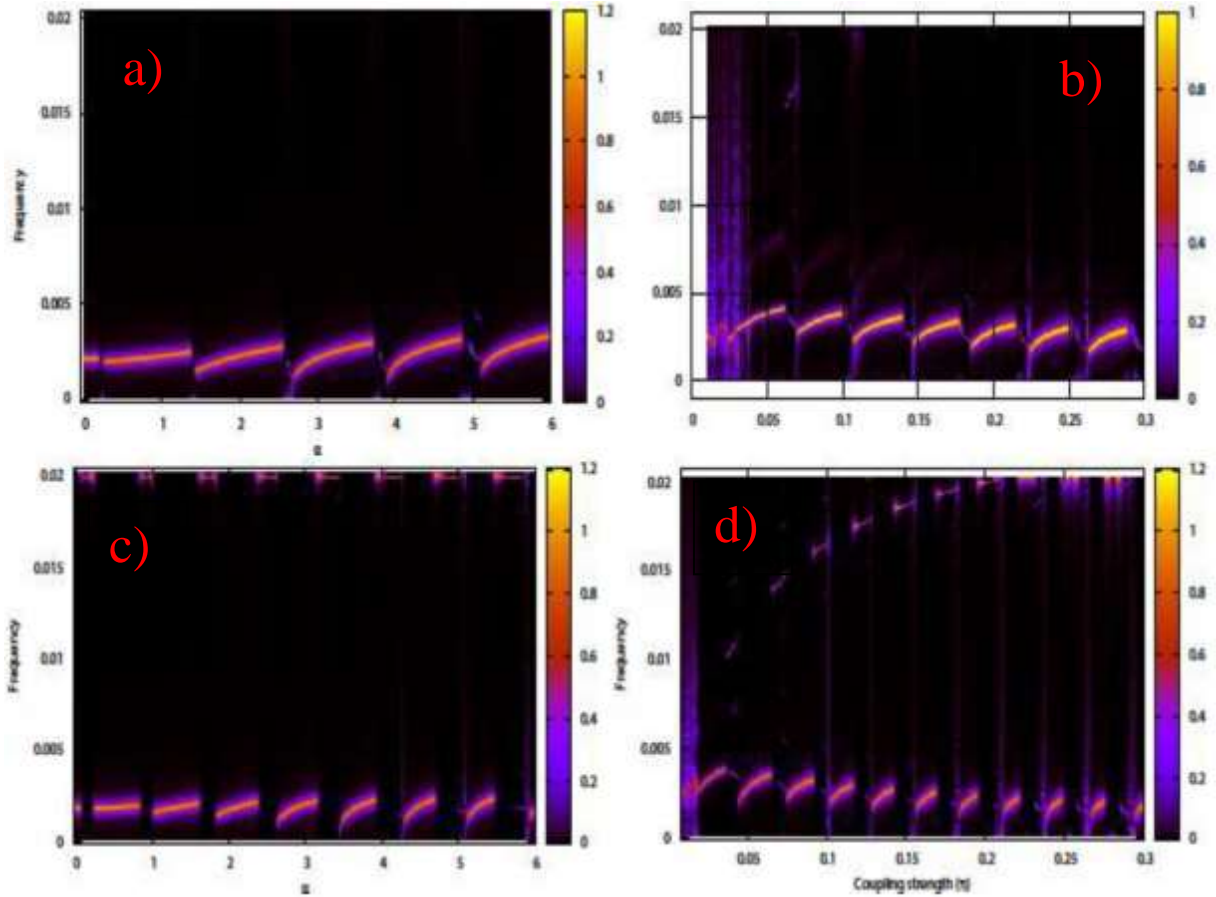

*Figure4.* Plot of frequency discontinuity vs α for different value of τ=14. in (a)and τ=50 in (c, ) ; dependence of frequency discontinuity and multistabilty between two consecutive frequency island on η for different value of τ in (b) and (d.)

We have systematically studied the various dynamical attractors (shown in figure 5) and the power spectral properties of the coupled laser system as a function of α and η for different values of τ [3, 5]. In a mutually coupled laser system, amplitude fluctuation in one laser leads to carrier density fluctuation, and through α phase fluctuation in the same laser [3].

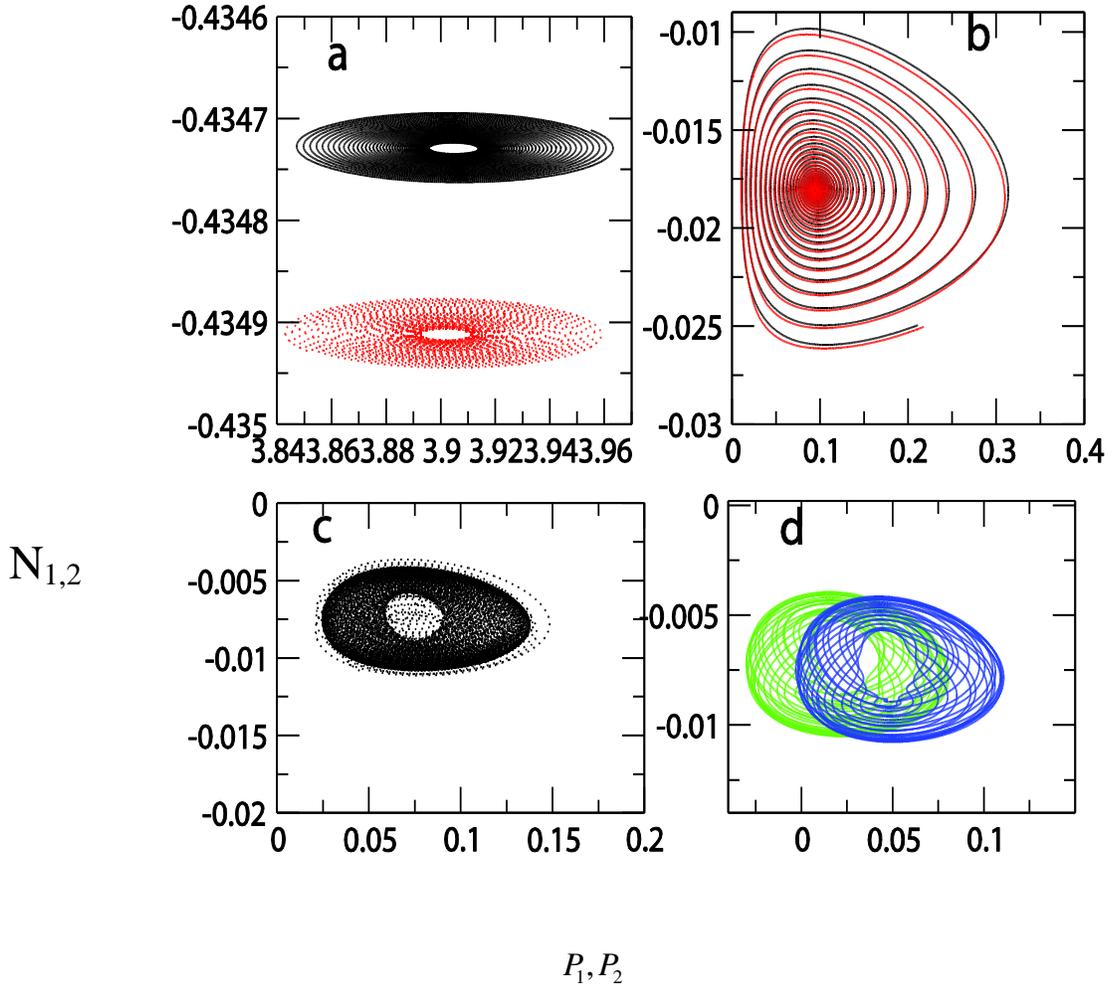

*Figure 5. Phase space plots of laser output powers $P_1$ and $P_2$ versus $N_1$ and $N_2$ for a fixed time delay of τ=14. in (a) anti-phase periodic state, (b) synchronized death state, (c) multi attractors , (d) co-existence of multi-attractors (multistability) in the gap between two consecutive frequency island in Fig.4.*

A change in the relative phase leads to an amplitude change in the second laser and an accompanying change in its carrier density [4] which in turn impacts the first. Therefore we explore the collective behaviour at different α and η, via spectral analysis [8]. The spectral diagrams (shown in Figure.4.) are obtained by combining the intensity spectra at different values of η and τ into a continuous intensity histogram. Instead of the continuous behavior of the frequencies (when τ=0) with changing α and η, we observe a" discretization" effect when some preferred frequencies appear. As illustrated in Figure. 4, the allowed values of the frequencies and their jumps are closely related to τ and strange bifurcation [2] near the transitions regimes. To show the influence of optical injection on α, Figure. 4 depict the variations of α at different η for fix time delay τ. Figure. 4(a) to 4(b) show the frequency variations and the multistability (Figure 5.) with in the ranges of frequency islands obtained numerically under different α vs η for fixed τ. Similar frequency quantization trend is found under different η with fixed τ and α. In Figure 6, we have observed frequency chaos as coupling strength induced modulation in alpha is very high.

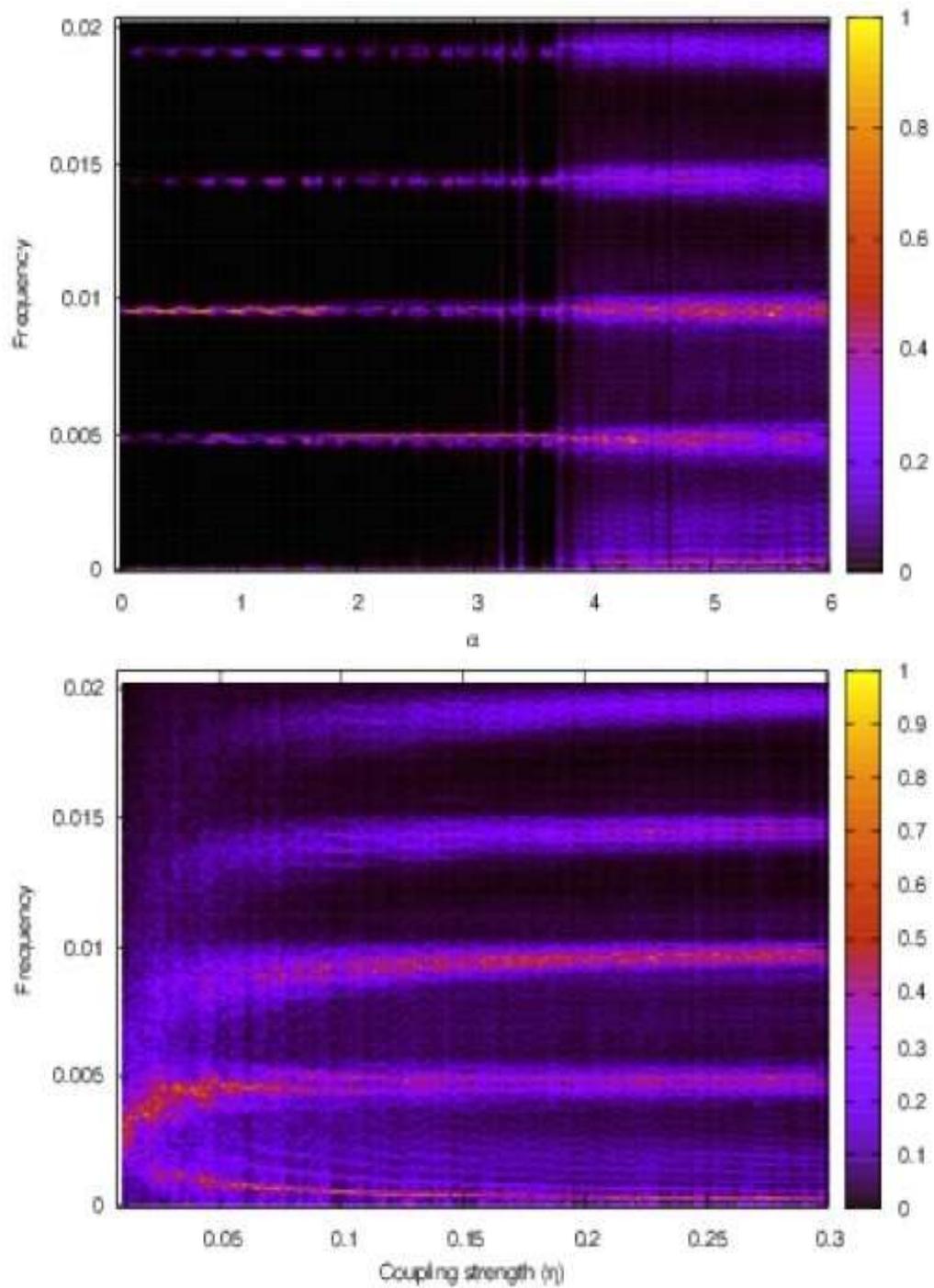

*Figure. 6.* Plot of frequency discontinuity vs α for the value of τ=100. in (top figure) ; dependence of frequency discontinuity and multistabilty between two consecutive frequency island on η for the same value of τ = 100 in (bottom figure).

From analytical and numerical analysis, we demonstrated that coherent self-sustained pulsations with different relative phase relationships between the electric field in the two lasers are possible (self-pulsating in-phase or out-of-phase super modes) for a wide range of parameters of the considered device ( as shown in figure 7). We have found two coherent regimes: stable CW in-phase and out-of-phase super-modes and in-phase and out-of-phase pulsating super-modes, where the intensity of the superposition of the two fields.. This could represent a promising result in view of the possibility of synchronizing a many-element array of pulsating lasers.

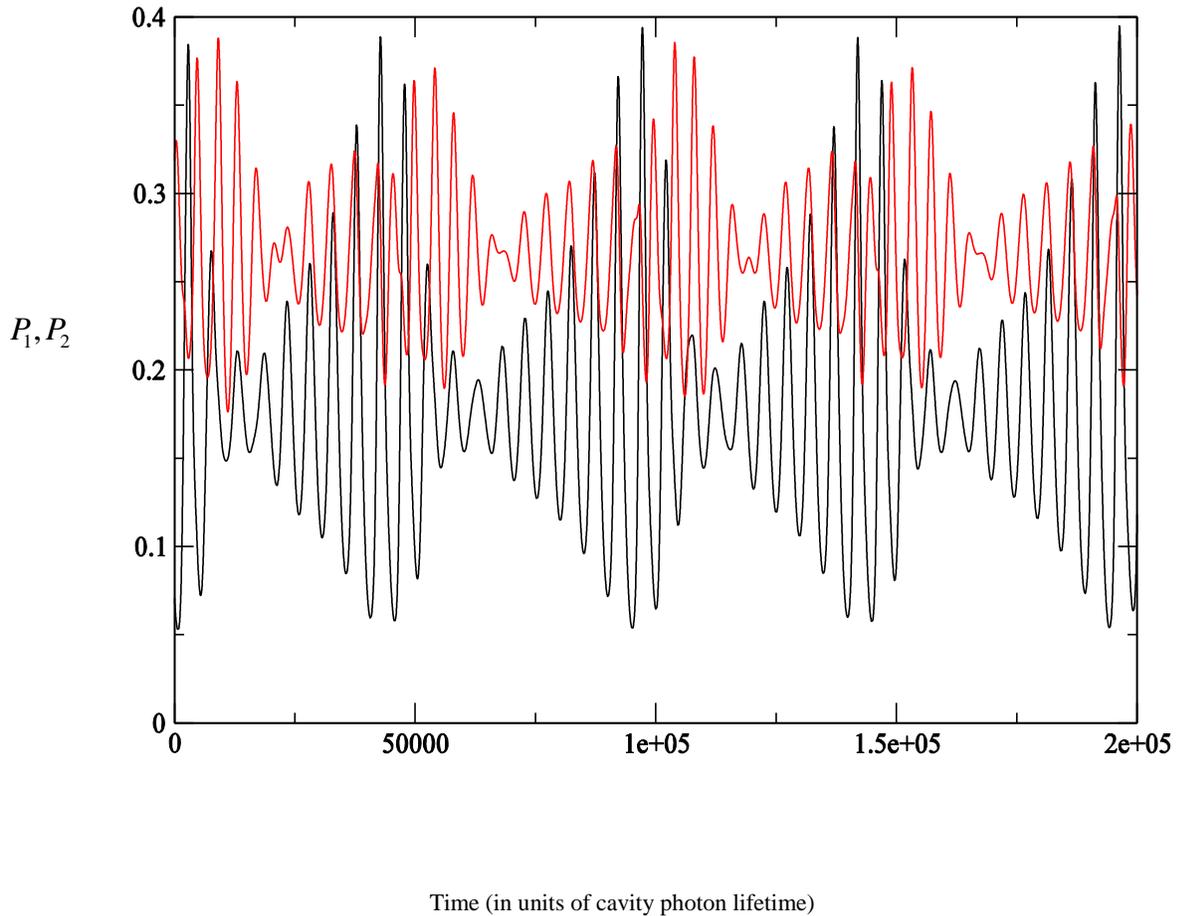

Time (in units of cavity photon lifetime)

Figure 7. . Plots of laser output powers P1 (black line) and P2 (red line) versus time (in units of cavity photon lifetime) for a fixed time delay $\tau=14$ in the same time units (left column) the out-of-phase periodic state in the transition regime; (right column) phase space plot of (left). The initial transients of $10^5$.

We have also observed periodic out-of-phase oscillation near and within the phase-flip transition regimes are shown in figure 8 [see figure 6]. Numerical analysis is based on the observation that the time-shifted correlation measure unveils the signature of coexistence of multiple attractors and enhancement of stability within anti-phase amplitude-death islands are shown in figure 5. Moreover, Passive mode-locking [14, 15] in the delay-coupled lasers system is also achieved in a small window of parameters near the phase transitions regimes without insertion any loss element as presented in Figure. 7. The pulses intensities are strongly modulated by a low-frequency envelope, so that they form pulse packages. This regime has been already identified in the dynamics of a laser with only optical feedback condition: it is called regular pulses packages (RPP) or Strange pulses [2].

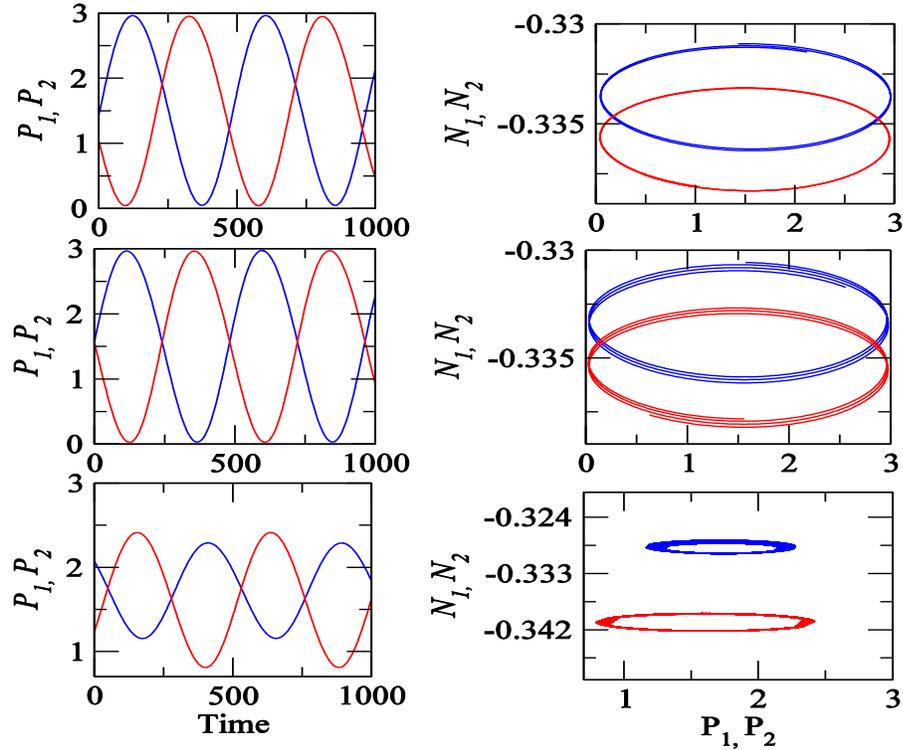

*Figure 8.* Plots of laser output powers P1 (black line) and P2 (red line) versus time (in units of cavity photon lifetime) for a fixed time delay τ=14 in the same time units (left column) the out-of-phase periodic state in the transition regime; (right column) phase space plot of (left). The initial transients of $10^5$.

Our finding distinguishes itself by its robustness against variations of system parameters, even in strongly coupled ensembles of oscillators.

## SUMMARY

In summary, we have demonstrated the effect of manipulating the phase-amplitude coupling factor α in mutually coupled lasers system with finite delay. We predict the occurrence of frequency discretization regimes and multistability which could be useful in controlling the chirp or pulse repetition rate of a photonic integrated compact device with the aid of phase control. We suggest that the dynamics are manipulated and controlled by changes in phase-amplitude coupling, and thus strong carrier dependence of the index on carrier density. Our observations provide a simple and low cost way to effectively control the RF dynamics of the delay-coupled semiconductor lasers system and here its uses as an optical clock, LIDAR. Instabilities need to be identified and studied, with a view to their suppression and exploitation in telecommunication networks.

## ACKNOWLEDGMENTS

PK and JMc would like to thank Science Foundation Ireland grant 12/IP/ 1258 for the financial support.

# REFERENCES

[1] Kumar, P., Prasad, A., & Ghosh, R., "Strange Bifurcation and Phase-locked Dynamics in mutually Coupled Diode laser Systems ", J. *Phys.B: At. Mol. Opt. Phys.*,Vol. **42**, pp. 145401 (2009).

[2] Kumar, P., Prasad, A., & Ghosh, R., "Stable phase-locking of an external-cavity diode laser subjected to external optical injection", J. *Phys.B: At. Mol. Opt. Phys.*,Vol. **41**, pp. 135402 (2008).

[3] Pal, V., Prasad, A., & Ghosh, R., "Optical Phase Dynamics in Mutually Coupled Laser System Exhibiting Power Synchronization", J. *Phys.B: At. Mol. Opt. Phys.*,Vol. **44**, pp. 235403 (2011).

[4] Mullane, M., P., & McInerney, John G., "Minimisation of the Linewidth Enhancement Factor in Compressively Strained Semiconductor Lasers", *IEEE Phot. Tech. Lett.,* Vol. **11**, pp. 776-778 (1999).

[5] Kumar, P., & Grillot, F., "Control of Dynamical Instability in Semiconductors Quantum Nanostructures Diode Lasers: Role of Phase-amplitude Coupling ", Eur. *Phys. J. Special Topic*,Vol. **222**, pp. 813-820 (2013).

[6] Henry, C. H., "Theory of the linewidth of semiconductor lasers," *IEEE J. Quantum Electron*., Vol. **18**, 259–264 (1982).

[7] Yu, Y., et al., "Influence of external optical feedback on the alpha factor of semiconductor lasers", *Opt. Lett.*, Vol. **38**, pp. 1781-1783 (2013).

[8] Pal, V., et al., "Semiconductor Laser Dynamics with two Filtered Optical Feedback", *IEEE J Quantum Electron*., Vol. **33**, pp. 1449-1454 (2013).

[9] Kelleher, et al., "Modified Relaxation Ocillation parameter in optically Injected Semiconductor Lasers", *J. Opt. Soc. Am. B.,*Vol. **29**, pp. 2249-2254(2012).

[10] Lazaridis, P., et al., "Time-bandwidth Product of Chirped Sech$^2$ Pulses : application to Phase-amplitude Coupling factor Measurement*", Opt. Lett.,* Vol. **20**, pp. 1160-1162 (1995).

[11] Hakki, P., B., et al., "Gain Spectra in GaAs Double-heterostucture Injection Lasers", *J. Appl. Phys.,* Vol. **46**, pp. 1299 (1975).

[12] Wang, C., et al., "Thermal Insensitive Determination of the Linewidth Broadening Factor in the Nanostructuresd Semiconductor Lasers Using Optical Injection Locking", *Scientfic Reports.,* Vol. **6**, pp. 27825 (2016).

[13] Murakami, A., et al., "Cavity Resonance Shift and Bandwidth Enhancement in Semiconductor Lasers with Strong Light Injection", *IEEE J Quantum Electron.,* Vol. **39**, pp. 1196 (2003).

[14] Kilen, I., Hader, J. and Moloney, J. V. & Koch, S. W., "Ultrafast nonequilibrium carrier dynamics in semiconductor laser mode locking", *Optica*, Vol. **1**, pp. 192-197 (2014).

[15] Keller, U. & Tropper, A. C. "Passively modelocked surface-emitting semiconductor lasers", *Phys. Rep*.,Vol. **429**, pp. 67-120 (2006).